# Three-terminal graphene single-electron transistor fabricated using feedback-controlled electroburning

Paweł Puczkarski[1], Pascal Gehring[1,*], Chit S. Lau[1], Junjie Liu[1], Arzhang Ardavan[2], Jamie H. Warner[1], G. Andrew D. Briggs[1] and Jan A. Mol[1]

[1]*Department of Materials, University of Oxford, 16 Parks Road, Oxford OX1 3PH, United Kingdom*

[2]*Clarendon Laboratory, Department of Physics, University of Oxford, Parks Road, Oxford OX1 3PU, United Kingdom*

[*]*pascal.gehring@materials.ox.ac.uk*

**Abstract**

We report room-temperature Coulomb blockade in a single layer graphene three-terminal single-electron transistor (SET) fabricated using feedback-controlled electroburning. The small separation between the side gate electrode and the graphene quantum dot results in a gate coupling up to 3 times larger compared to the value found for the back gate electrode. This allows for an effective tuning between the conductive and Coulomb blocked state using a small side gate voltage of about 1V. The technique can potentially be used in the future to fabricate all-graphene based room temperature single-electron transistors or three terminal single molecule transistors with enhanced gate coupling.

Due to its 2D character and high charge carrier mobility graphene is one of the most promising materials for future electronics.[1] Progress in chemical vapour deposition growth allows for the fabrication of graphene foils with sizes up to 750 mm or more.[2] Graphene can be patterned with dimensions down to a few nanometres using standard lithography techniques,[3,4] which opens up the possibility of scalable fabrication of graphene based nanoelectronics.[5] Graphene quantum dots have been successfully fabricated by means of lithography[3] and electroburning techniques[6,7]. The latter allows the formation of single electron transistors (SETs) with addition energies up to 1.6 eV, enabling room temperature operation.[6] Side gate electrodes in close proximity to the quantum dot give reasonably strong gate coupling.[3] Graphene nano-gaps formed by electroburning offer a promising platform for contacting single molecule transistors.[8,9] Strong gate coupling in single-molecule devices has been achieved by using gate dielectrics with a high dielectric constant, by fabricating devices with ultra-thin gate dielectrics.[10] Until now all studies on electroburned graphene devices used a global $SiO_2$/Si back gate system with oxide thicknesses of several hundreds of nm which results in a comparatively low gate coupling.[6] In this letter we report the observation of Coulomb blockade in a graphene SET where the source, drain and side gate electrodes are fabricated using feedback-controlled electroburning. If the details of the breakdown of graphene could be understood, this fabrication technique will be a further step towards scalable graphene electronics.

We use a combination of electron beam lithography and feedback-controlled electroburning[11,12] to fabricate three-terminal graphene devices. CVD-grown single layer graphene (SLG) is transferred onto a Si/$SiO_2$ chip that is pre-patterned with gold electrodes. After transferring the SLG, it is patterned into a Y-shaped geometry (see Figure 1) by exposing a negative resist using electron beam lithography followed by oxygen plasma etching.

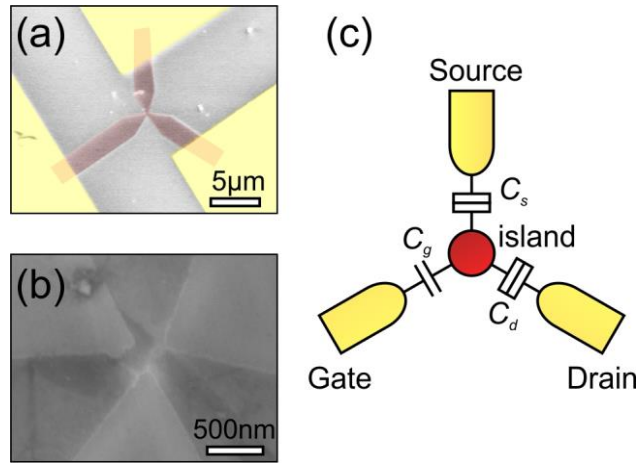

**Figure 1.** (a) Scanning electron microscopy image of a typical three-terminal graphene device. Graphene and gold contacts are false-coloured in red and yellow, respectively. (b) Magnification of a graphene device. A gap between source/drain and gate electrode can be observed. (c) Schematic drawing of the all-graphene single electron transistor (SET) depicted in (b).

The Y-shaped geometry is designed such that the narrowest parts of the three terminals are interconnected at the centre of the device. As a result, the current density during electroburning is highest in these regions, so that the nano-gaps will form at the constrictions.[11]

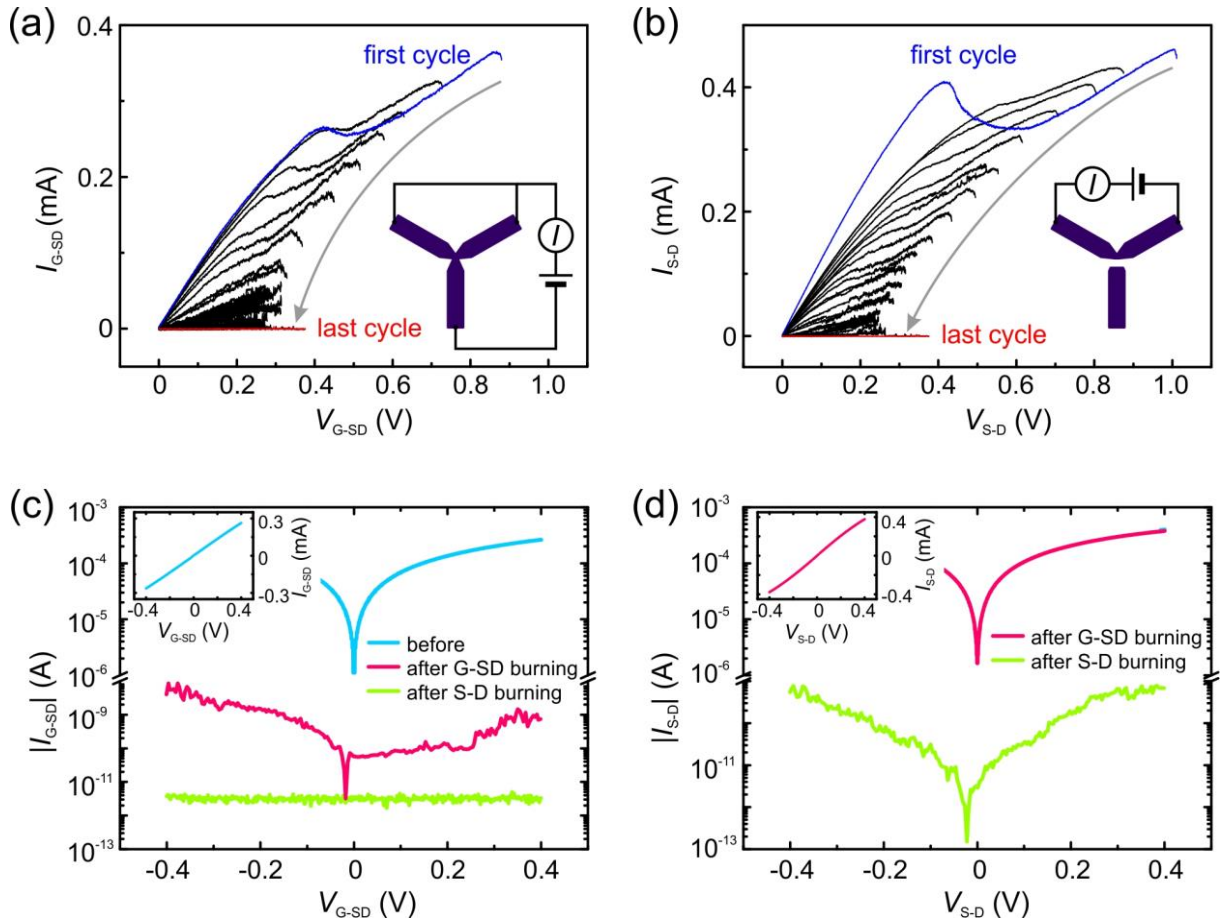

**Figure 2.** I-V traces recorded during the feedback controlled electroburning of the (a) G-SD and the (b) S-D gap. The traces of the as-prepared and the fully broken device are shown in blue and red, respectively. The insets depict the connections used for electroburning. (c) Comparison of G-SD *I-V* traces before electroburning (blue), after opening a G-SD gap (pink) and after burning a S-D gap (green). Inset: G-SD *I-V* trace before electroburning on a linear scale. (d) Comparison of S-D *I-V* traces after opening a G-SD gap and before electroburning of the S-D electrodes (pink) and after burning a S-D gap (green). Inset: S-D *I-V* trace before electroburning on a linear scale.

The feedback-controlled electroburning is performed in air at room temperature using an automated probe station. The fact that the devices are patterned on a regular grid allows us to test and burn 480 devices in about 64 hours using the automated probe station. The fabrication process consists of two electroburning steps. In the first electroburning step a gap is formed between the source/drain channel and the gate electrode. The source (S) and drain (D) terminals are shorted and a voltage is applied between the source/drain (SD) terminal, and the gate (G) terminal (see inset in Figure 2a). As the voltage is increased the current is monitored with a 5 kHz sampling rate. When the feedback condition, which is set at a drop $\Delta I_{set}$ of the current within the past 15 mV, is met the voltage is ramped back to zero. After each burning cycle the resistance between the gate and source and drain terminals is measured and the process is repeated until the low-bias resistance exceeds a threshold resistance $R_{set}$. A typical evolution of the current-voltage (*I-V*) traces is shown in Figure 2a. To prevent the device breaking at the initial voltage ramps, the feedback conditions are adjusted for each burning cycle depending on the threshold voltage $V_{th}$ at which the previous drop occurred. After a gap is formed between the gate electrode and the source/drain channel, we use a second electroburning step to form a gap between the source and drain electrodes. In this second electroburning step the gate terminal is disconnected and left floating, while a voltage is applied between the source and drain terminals (see inset in Figure 2b). The feedback-controlled electroburning process is repeated until the resistance between the source and drain electrodes exceeds $R_{set}$ (see Figure 2b).

To characterise the size of the gaps formed during the breakdown of graphene *I-V* traces were recorded before and after electroburning (see Figure 2c and d). The shape of the source/drain and gate *I-V* curves changes from linear Ohmic to exponential tunnelling behaviour upon electroburning, and the low bias resistance increases by approximately five orders of magnitude. We estimate the width $d$ and the average height $\phi = \frac{\phi_L + \phi_R}{2}$ of the tunnelling barrier for electrons, where $\phi_L$ and $\phi_R$ are the work functions of the left and right lead, respectively, by fitting the data to the Simmons model.[11,13] After the first electroburning step the average width of the gap between the source/drain channel and the gate electrode (G-SD gap) measured for 103 devices was $d_{G-SD} = 1.6 \pm 0.7$ nm with an average barrier height of $\phi_{G-SD} = 0.22 \pm 0.12$ eV. After the second electroburning step the average size of the gap between the source and drain electrode (S-D gap) measured for 71 devices was $d_{S-D} = 1.8 \pm 0.8$ nm with an average barrier height of $\phi_{S-D} = 0.23 \pm 0.10$ eV. These values are small compared to the work function of bulk graphite, however, similar values have previously been observed in studies of electroburnt single layer[11] and few layer[8] graphene devices. We did not observe a correlation between the size found for the G-SD and S-D gap.

After the second electroburning step (S-D burning) we no longer observe a tunnel current between the gate and the source/drain terminals (see Fig 2c). This indicates that during the formation of the source-drain gap, the gap between the gate and the source-drain channel is widened further. We attribute the widening of the G-SD gap to the removal of carbon islands remaining after the first electroburning step or to further temperature activated oxidation of carbon at the edge of the side gate electrode.

To demonstrate that the third electrode can be used for electrical gating we investigated devices where the S-D gap did not break completely or which contained small carbon islands inside the S-D gap, both indicated by a linear *I-V* characteristic around zero bias (see Figure 4b below) which cannot be fitted using the Simmons model. In these devices charge inhomogeneity in the $SiO_2$ which result in charge puddles or any kind of edge disorder can give rise to charge localisation inside the constriction.[6,7,14] We performed gate dependent measurements of 66 devices selected in this way and identified 10 devices with a pronounced gate effect. Figure 3a shows the measured source-drain current *I* as a function of the bias $V_{bias}$ applied to the drain electrode and the voltage $V_{side}$ applied between the side gate and the source electrode at *T* = 2.2 K for the device of Figure 2. A Coulomb diamond with a region of single electron tunnelling at $V_{side}$ = 1.2 V indicated by the dotted lines can be observed. From the positive and negative slope of the edges of the conductive region we can extract $C_s = 14 C_{g,side}$ and $C_d = 20 C_{g,side}$,[15] where $C_s$, $C_d$ and $C_{g,side}$ are the capacitances between the quantum dot and the source, drain and side gate electrode, respectively. From relative capacitive coupling strengths we can infer that the graphene island is well located between source and drain electrode ($C_s \sim C_d$), with a slightly higher coupling to the drain, whereas the distance between the island and the side gate is found to be larger ($C_{g,side} < C_s$, $C_d$). From the capacitances we can also calculate the lever arm $\alpha_{side} = C_{g,side} / (C_{g,side} + C_s + C_d) = 0.03$ that quantifies the shift of the electrochemical potential of the quantum dot as a function of the gate voltage. While the value of the lever arm observed in our device is smaller than 0.1 typically found for 3 nm thin $Al_2O_3$ oxide layers,[10] it is approximately 2-3 times larger than the values $\alpha_{back}$ measured for our $SiO_2$ (300nm)/Si back gate electrode and slightly higher than that for lithographically defined graphene side gate structures where $\alpha_{side}/\alpha_{back} \leq 1.15$.[3] In our experiments we observe an increasing current at high negative and positive side gate voltages, which we attribute to the leakage tunnel current through the barrier between the side gate and the source/drain channel. The G-SD barrier becomes transparent at a breakdown voltage of approximately $|V_{side}| > 2.5$ V where $|I_{leak}| > 0.1$ nA, which limits the gate-induced shift of the electrochemical potential of the quantum dot to $\alpha_{side}V_{side} = \pm 75$ meV.

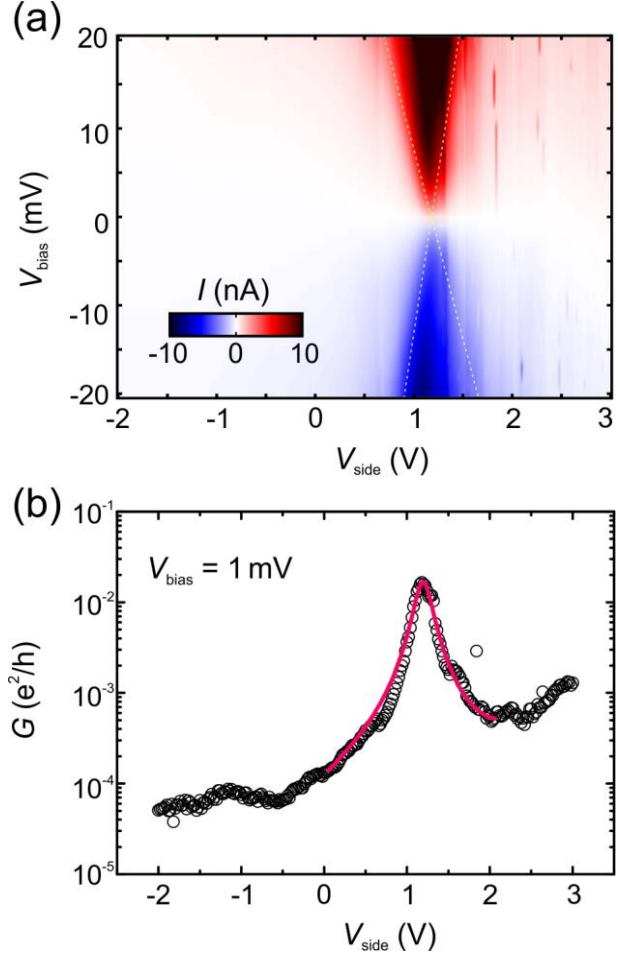

**Figure 3.** (a) Map of the source-drain current as a function of the bias and side-gate voltage at $T = 2.2$K. The position of the Coulomb diamond is highlighted by the dashed yellow lines. (b) Numerical conductance d$I$/d$V$ as a function of side-gate voltage at $V_{bias} = 1$mV. The Coulomb peak is fitted (pink curve) using Eq. 1.

The asymmetry in the island-lead capacitances suggests that the island is not exactly centred between the two leads. This will also influence the tunnel coupling to the leads which depends exponentially on the separation between the island and the electrodes. When the electron temperature $T$ of the leads is sufficiently low, such that $k_B T \ll h\Gamma$, where $\Gamma = \Gamma^S + \Gamma^D$ is the sum of the tunnel rates to source and drain, the Coulomb peak is described by the Breit-Wigner form:[16]

$$G = \frac{2e^2}{h}\left(\frac{1}{\Gamma^S} + \frac{1}{\Gamma^D}\right)^{-1} \frac{h^2 \Gamma}{\alpha^2(V_0 - V_{\text{side}})^2 + (h\Gamma/2)^2}, \qquad (1)$$

where $V_0$ is the centre of the peak. From the full width at half maximum of the Coulomb peak shown in Fig. 3b we obtain a coupling strength of $h\Gamma = 6$ meV which is an order of magnitude larger than $k_B T \approx 0.2$ meV at 2.2 K. From this we can conclude that, in this device at 2.2 K, the width of the Coulomb peak is dominated by lifetime broadening. The asymmetry of the tunnel coupling strength to source and drain electrode can be investigated by studying the asymmetry of the *I-V* curves. Inside the conductive region around $V_{\text{side}} = 1.2$ V we observe a current ratio $I(V_{sd})/I(-V_{sd}) \approx 0.55$, which we attribute to the asymmetry of the source and drain tunnel barriers. In the sequential transport regime, the tunnel current *I* through a single level is given by:[17]

$$I = e \frac{2\Gamma^{\text{in}} \Gamma^{\text{out}}}{2\Gamma^{\text{in}} + \Gamma^{\text{out}}}, \qquad (2)$$

where $2\Gamma^{\text{in}}$ accounts for the fact that there are two transport channels to tunnel into an empty spin-degenerate level, namely spin up and spin down, whereas there is only one transport channel to tunnel out of the level. For $V_{sd} > 0$ electrons tunnel from the source into the quantum dot and out to the drain, so that $\Gamma^{\text{in}} = \Gamma^S$ and $\Gamma^{\text{out}} = \Gamma^D$, while for $V_{sd} < 0$ electrons tunnel from the drain into the dot and out to the source, and $\Gamma^{\text{in}} = \Gamma^D$ and $\Gamma^{\text{out}} = \Gamma^S$. As a result, the current ratio

$$\frac{I(V_{sd})}{I(-V_{sd})} = \frac{\beta + 2}{2\beta + 1}, \qquad (3)$$

can be expressed as a function of the ratio $\beta = \Gamma^S/\Gamma^D$ between the coupling strengths to the source and drain electrodes. From the current ratio 0.55 we deduce that $h\Gamma^S = 0.4$ meV and $h\Gamma^D = 5.6$ meV, consistent with the aforementioned difference in capacitive coupling strengths to the source and drain electrode.

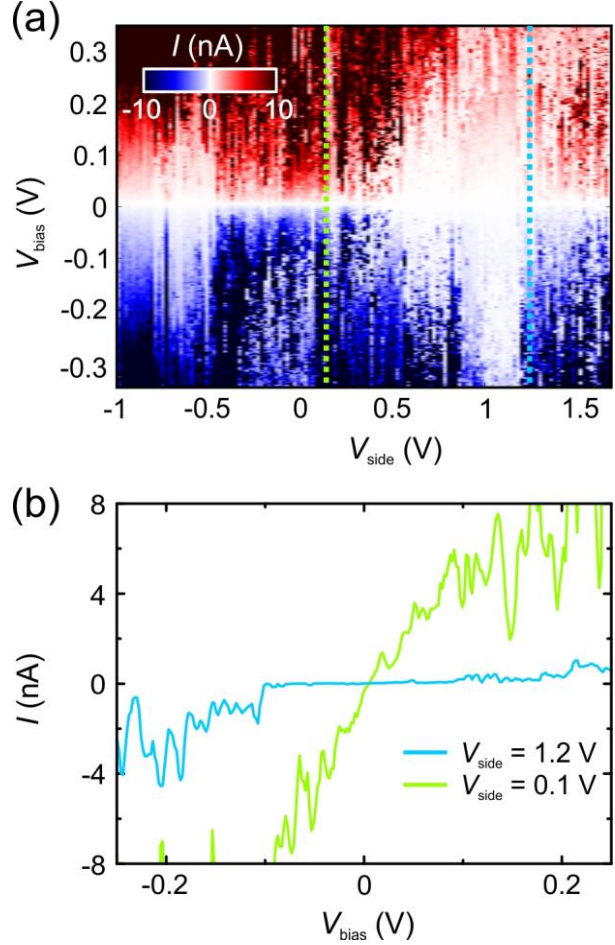

**Figure 4.** (a) Map of the source-drain current as a function of the bias and side-gate voltage at room temperature. Typical *I-V* traces inside the conductive (dashed green line) and the blocked (dashed blue line) region are depicted in (b).

Finally we investigate room-temperature operation of the all-graphene SET. Figure 4a shows a stability diagram recorded at $T = 300$ K and under ambient pressure. Regions of high and low current can be observed, which are further emphasised by line cuts extracted at two different side gate voltages (see Figure 4b). From this data a gate coupling factor of $\alpha = 0.26$ can be extracted, which increased by an order of magnitude at room temperature compared to its value at 2.2 K. We attribute this to small charge rearrangements in the side gate which reduce the effective G-SD gap when warming up the sample. These charge rearrangements also shift the positon of the Coulomb peak. The enhanced gate coupling allowed us to observe a second charge state at about $V_{side} = -0.6$ V. From the width and height of the Coulomb diamond an addition energy of $E_{add} \approx 300$ meV can be extracted. We will use the value of the addition energy to estimate the size of the dot following Ref 6. A lower limit for the size can be calculated by assuming that the quantum dot level-spacing $\Delta E$ is small compared to the charging energy $E_c$, so that $E_c \approx E_{add}$. From $E_c = e^2/2C$ we obtain a total capacitance of the quantum dot of $C = 0.27$ aF. This corresponds to the self-capacitance of a circular disk at the interface between air and SiO$_2$ with a diameter $d = C / (2\varepsilon_0(\varepsilon_{oxide} + \varepsilon_{air})) = 3.1$ nm, using the vacuum permittivity $\varepsilon_0$ and the relative dielectric constants $\varepsilon_{oxide} = 3.9$ of SiO$_2$ and $\varepsilon_{air} = 1.0$ of air. A second estimation can be made by assuming that the level-spacing $\Delta E$ dominates the addition energy. If we assume that the electrons in the quantum dot are confined in a two-dimensional square-well potential, we can estimate the diameter of the monolayer graphene island by $d = \pi \hbar v_F / E_c = 6.9$ nm, where $v_F = 1 \times 10^6$ m/s is the Fermi velocity of graphene and $\hbar$ is the reduced Planck constant.[6] Both estimations indicate that the size of the graphene island is of the order of several nanometres.

To conclude, we fabricated all-graphene three terminal single electron transistors with a high side gate coupling using an electroburning technique. Due to their high addition energy and gate coupling strength the SETs can be operated at room temperature with a side gate voltage < 1V. Single molecules can be effectively coupled to graphene electrodes using suitable anchor groups.[8,9] This idea can be directly applied to our devices by choosing empty gaps with a gap size of 1-2 nm to fabricate three terminal single molecular transistors with enhanced gate coupling.

**Acknowledgements**

We thank the Royal Society for a Newton International Fellowship for J. A. M. and a University Research Fellowship for J. H. W., and the Agency for Science Technology and Research (A*STAR) for a studentship for C. S. L. We acknowledge C. Salter and the Oxford University Materials Characterisation Service for providing electron microscopy images. This work is supported by Oxford Martin School, EPSRC grants EP/J015067/1. This project/publication was made possible through the support of a grant from Templeton World Charity Foundation. The opinions expressed in this publication are those of the author(s) and do not necessarily reflect the views of Templeton World Charity Foundation.